\documentclass[aps, prb, 10pt, twocolumn, reprint, superscriptaddress, amsmath, amssymb]{revtex4-2} 

\usepackage{graphicx}
\usepackage{dcolumn}
\usepackage{bm}

\usepackage{float}
\usepackage{hyperref}

\usepackage[usenames]{xcolor}

\usepackage{multirow}
\usepackage{threeparttable}

\begin{document}
\newfloat{scheme}{hhtbp}{lof}
\floatname{scheme}{\small SCH.~}


\title{Microscopic Theory for the Incoherent Resonant and Coherent Off-Resonant Optical Response of Tellurium}

\author{Sven C. Liebscher}

\author{Maria K. Hagen}%
\affiliation{ 
Department of Physics and Material Sciences Center, Philipps-University Marburg, Renthof 5, 35032 Marburg, Germany
}%

\author{J{\"o}rg Hader}

\author{Jerome V. Moloney}
\affiliation{%
Wyant College of Optical Sciences, University of Arizona, Tucson, Arizona 85721, USA
}%

\author{Stephan W. Koch}
\affiliation{ 
Department of Physics and Material Sciences Center, Philipps-University Marburg, Renthof 5, 35032 Marburg, Germany
}%

\date{\today}%

\begin{abstract}
An $\it{ab \,\, initio}$ based fully microscopic approach is applied to study the nonlinear optical response of bulk Tellurium. The structural and electronic properties are calculated from first principles using the shLDA-1/2 method within density functional theory. The resulting bandstructure and dipole matrix elements serve as input for the quantum mechanical evaluation of the anisotropic linear optical absorption spectra yielding results in excellent agreement with published experimental data. Assuming quasi-equilibrium carrier distributions in the conduction and valence bands, absorption/gain and spontaneous emission spectra are computed from the semiconductor Bloch and luminescence equations. For ultrafast intense off-resonant excitation, the generation of high-harmonics is evaluated and the emission spectra are calculated for samples of different thickness.

\end{abstract}


\maketitle


\section{Introduction}

Elemental Tellurium is known to have a wide variety of unusual optical characteristics. It is the only elemental semiconductor with a direct bandgap in the technically interesting mid infrared wavelength range near $3.8\mu\mathrm{m}$. Furthermore, Te is considered to have exceptional non-linear optical properties\cite{fee1970,Berezovskii1972} due to its chiral structure where the atoms form helical chains. Relatively few studies of the optical properties of Te have been published so far. Reflectivity and absorption spectra and their temperature dependent variations have been analyzed in Refs.\onlinecite{Sobolev63,Stuke64,Loferski54,Tutihasi1969} and references therein. These papers report a strong polarization dependence of the optical response, reflecting the uniaxial nature of the Te crystal. Bulk crystals of Te exhibit large refractive indices with a prominent difference between the ordinary and extraordinary directions (about 4.9 and 6.3 near the bandgap)\cite{Caldwell59}. 

The strongly directional crystal structure also leads to prominent optical nonlinearities. An exceptionally large nonlinear coefficient was confirmed by a phase-matched harmonic generation measurement on an elemental Te crystal in Ref.\onlinecite{Patel65}. Furthermore, the chiral structure has been shown to lead to gyroscopic nonlinear optical responses depending on the helicity of the light (see e.g. Ref.\onlinecite{Tsirkin18} and references therein). Measurements of the photoluminescence from bulk Te crystals have been reported for cryogenic temperatures in Ref.\onlinecite{Benoit65} and for room temperature in Ref.\onlinecite{Choi19}. These publications also document indications of stimulated emission and lasing as well as strong second- and third-order harmonic generation.

To complement and extend the earlier investigations, we present in this paper a comprehensive analysis of the nonlinear optical properties of bulk Te. For this purpose, we performed a systematic microscopic study of its resonant incoherent and off-resonant coherent properties. We employ an $\it{ab \,\, initio}$ based approach where we use Density Functional Theory (DFT) together with the shell Local Density Approximation-1/2 (shLDA-1/2) method to obtain accurate structural and electronic parameters. We evaluate the dispersion of the energetically highest valence and the lowest conduction bands and determine the relevant dipole and Coulomb interaction matrix elements. 

Using these results as input for the semiconductor Bloch equations (SBE)\cite{lindberg1988}, we first evaluate the Te absorption spectra for different excitation conditions. Our results show excellent agreement with published experimental data. Assuming quasi equilibrium carrier populations in the relevant valence and conduction bands, we compute the transition from absorption to optical gain. The corresponding luminescence spectra are evaluated using the semiconductor luminescence equations (SLE)\cite{sle}. Both, gain and luminescence exhibit strong dependence on the light polarization direction. 

For strongly off-resonant excitation, we investigate the generation of high harmonics in a wide spectral range extending far above the fundamental Te bandgap. Currently, high harmonic generation (HHG) in semiconductors after excitation with short high-intensity pulses is a field of active research\cite{ghimire2011observation,  luu2015extreme, yoshikawa2017high, vampa2015linking, Hohenleutner2015,  ndabashimiye2016solid, liu2017high, you2017anisotropic, xia2018nonlinear, kemper2013theoretical, Vampa2014, hawkins2015effect, higuchi2014strong, tamaya2016diabatic, wu2015high, ghimire2012generation}. Microscopically, semiconductor HHG can be related to the nonequilibrium dynamics of the induced electron-hole excitations, including interband polarizations and intraband currents probing the conduction and valence bandstructure in the entire Brillouin (BZ) zone. To analyze these effects, we use our DFT results as structural input for the SBE and compute HHG spectra for different excitation conditions. Besides local evaluations of HHG spectra, we also study the effects of different sample thicknesses performing calculations which explicitly include field propagation effects.

This paper is organized as follows: In Sec. 2, we give an overview of our DFT approach and discuss the resulting bandstructure and the relevant dipole matrix elements for bulk Te. Section 3 summarizes our calculations for optical absorption, gain, and photo luminescence, whereas Sec. 4 is devoted to the modeling of HHG in Te for different excitation conditions and sample lengths. A short summary and outlook in Sec. 5 concludes our presentation. 

\section{Electronic Structure Calculations}
\subsection{Computational Details}
In our approach to construct the electronic structure for bulk Te, we use the Vienna Ab initio Simulation Package\cite{Kresse1993, Kresse1994, Kresse1996, Kresse1996a} (VASP) version 5.4.4 which implements the Projector-Augmented Wave (PAW) method\cite{Kresse1999, Blochl1994}.
Starting from the symmetry group of right-handed Tellurium $P3_121-D^4_3$, the crystal structure was relaxed using the Generalized Gradient Approximation (GGA) by Perdew, Burke and Ernzerhof (PBE)\cite{Perdew1996} for the exchange-correlation energy.
A $\Gamma$-centered Monkhorst-Pack\cite{Monkhorst1976} grid of $15 \times 15 \times 15$ $k$-points and a plane wave basis-set cutoff energy of $500 \, \text{eV}$ was used.
The cell volume, cell shape and ion positions were optimized using the conjugate gradient algorithm.
The convergence criteria were set to $10^{-9} \, \text{eV}$ for electronic minimization and $3 \cdot 10^{-4} \, \text{eV}/\text{\AA}$ for the forces acting on the ions.\\

After relaxation, the PAW pseudopotential for Te was modified according to the shLDA-1/2 method as proposed by Xue et al.\cite{Xue2018}.
This method is based on the LDA-1/2\cite{Ferreira2008, Ferreira2011a} method, which aims to avoid the underestimation of bandgaps with a GGA by correcting for the self-interaction of a localized hole in the valence band by adding a so-called self-energy potential to the pseudopotential.
Based on Slater's half-occupation technique\cite{Slater1972}, the self-energy potential is found by subtracting the potential of the half-ionized atom from the unionized atom.
Since this self-energy potential is added to every atom, it has to be trimmed to avoid divergent contributions. 
In the LDA-1/2 method, this is achieved with a spherical trimming function
\begin{equation}
 \Theta(r) = 
 \begin{cases}
  \left[1-\left(\frac{r}{r_\text{cut}}\right)^n\right]^3 & r \leq r_\text{cut} \\
  0 & r > r_\text{cut}
 \end{cases}\quad ,
\end{equation}
in which the cutoff radius has to be determined variationally with the condition that the resulting bandgap is maximized.
In the shLDA-1/2 method, the trimming function is replaced by a spherical shell
\begin{equation}
 \Theta(r) = 
 \begin{cases}
  \left[1-\left(\frac{r}{r_\text{out}}\right)^m\right]^3 \frac{1+\text{tanh}[n(r-r_\text{in})]}{2} & r \leq r_\text{out} \\
  0 & r > r_\text{out}
 \end{cases} ,
\end{equation}
which is more suitable for crystals where the charge is not centered around the atom cores, but lies between two atoms.
In this case, in addition to the outer cutoff radius $r_\text{out}$ an inner cutoff radius $r_\text{in}$ has to be determined by the same method as before, keeping the outer cutoff radius constant.
The self-energy corrected pseudopotentials for different cutoff radii have been constructed and the optimal cutoff radius determined by fitting a quadratic function of the cutoff radius to the resulting bandgaps and finding the maximum.
The corresponding DFT calculations used the same computational parameters as the relaxation, however, the crystal structure was kept constant and spin-orbit coupling was included.\\

In a third set of calculations, the band structure and dipole matrix elements were determined. 
To this end, the charge-density of the self-consistently calculated ground-state obtained with the constructed pseudopotential was read in and kept constant.
The $k$-points were chosen along high symmetry lines in the Brillouin zone and the number of bands was increased, since a significant amount of empty conduction bands is needed for the optical routines of the VASP program that calculate the dielectric properties\cite{gajdos2006}.
\subsection{Bandstructure and Dipole Matrix Elements}
\label{dftres}
\begin{table}
\caption{Comparison of structural and electronic parameters from \textit{ab-initio} DFT calculations using the shLDA-1/2 method with experimental results.}\label{tab:dftres}
\begin{ruledtabular}
\begin{tabular}{lccccc}
 & \multicolumn{3}{c}{Structural parameters} & \multicolumn{2}{c}{Electronic Properties} \\
 & $a$ & $c$ & $u$ & $E_g$ & $E_{\text{LH-HH}}$ \\
\hline
DFT & $4.51 \, \text{\AA}$ & $5.96 \, \text{\AA}$ & $0.27$ & $0.323 \, \text{eV}$ & $0.111 \, \text{eV}$ \\
Exp.\cite{Adenis1989, Anzin1977, Caldwell1959} & $4.46 \, \text{\AA}$ & $5.92 \, \text{\AA}$ & $0.267$ & $0.33 \, \text{eV}$ & $0.112 \, \text{eV}$
\end{tabular}
\end{ruledtabular}
\end{table}
\begin{figure}[htbp]
    \includegraphics{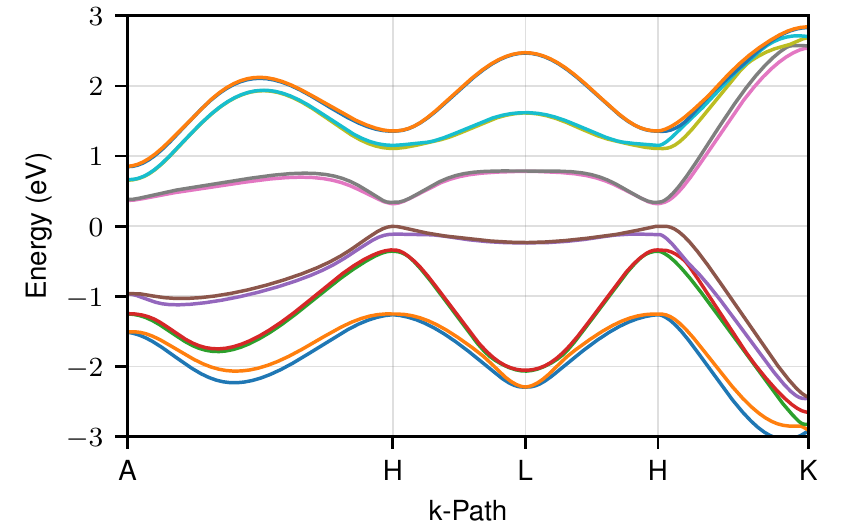}
    \caption{Lowest six electron bands and highest six hole bands of Tellurium calculated with the shLDA-1/2 method.}
    \label{bstr}
\end{figure}
\begin{figure*}[ht]
    \includegraphics[width=.89\textwidth]{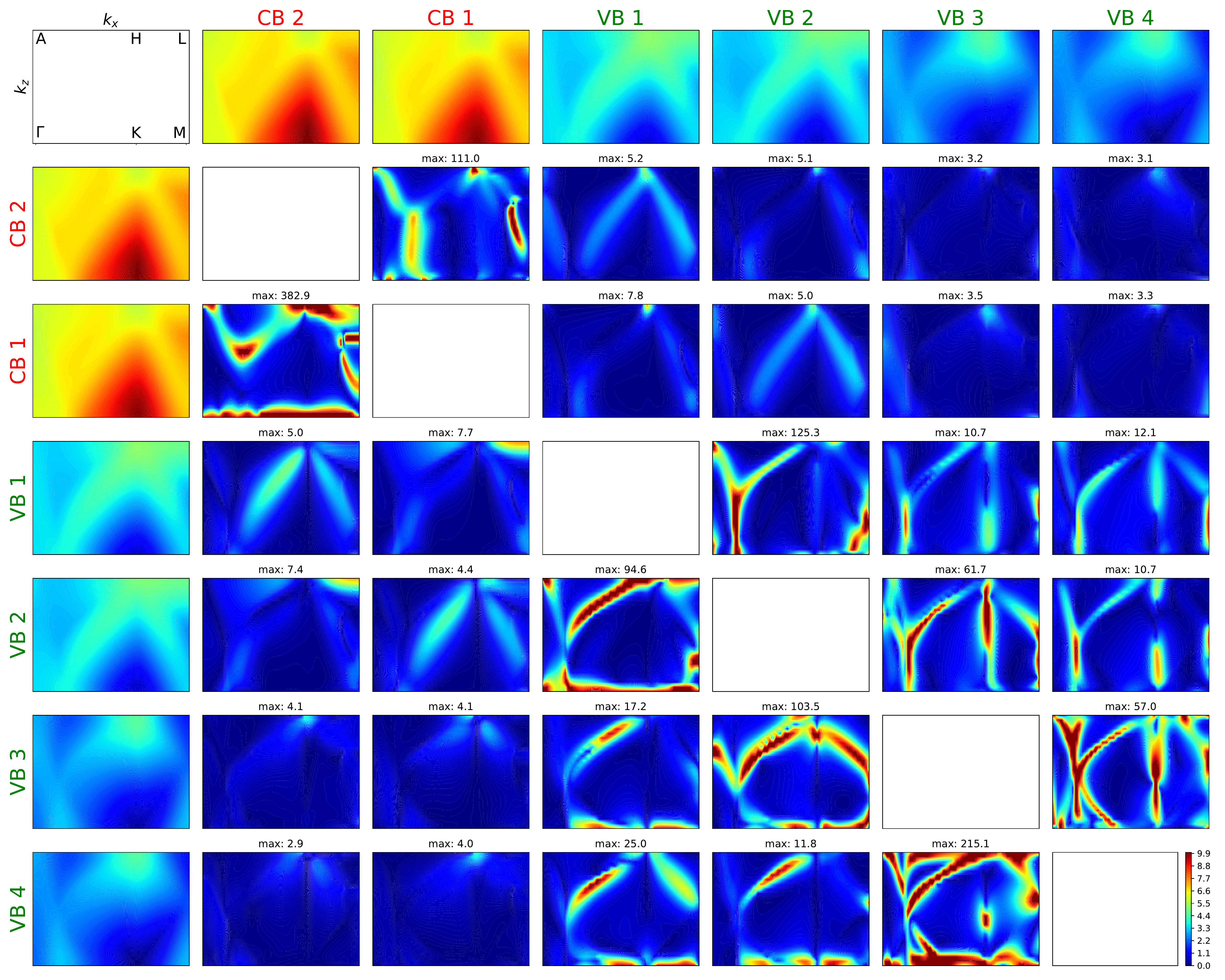}
    \caption{Dipole transition matrix elements between bands in a 2d plane of the 1. BZ spanned by the $\Gamma$-,M-,L- and A-points. The first row and first column show the band energies, while the inner plots show the dipole strengths. The dipole in a particular cell corresponds to the band combination given by the bands in the respective row and column. In the bottom left triangle, the dipoles for $E \parallel c$ direction are given, in the top right triangle, the dipoles for $E \perp c$ direction are given. The color bar in the bottom right cell pertains to all dipoles. Values higher than 10 are not distinguished in the color plot. The 'max' value above each dipole plot indicates the maximum value of the respective dipole at any point in the plane. }
    \label{dips}
\end{figure*}
The results of the structural relaxation can be found in the first three columns of Table \ref{tab:dftres}, where $a$ and $c$ are the lattice constants and $u$ is the parameter that determines the position of the atoms in the plane perpendicular to the helical chains.
Comparison to experimental values shows that both lattice constants are slightly overestimated.
For the construction of the self-energy corrected pseudopotential, the optimized inner and outer cutoff radii were determined as $1.328 \, \text{\AA}$ and $3.395 \, \text{\AA}$, respectively.
The resulting direct bandgap at the H-point, $E_g$, and  splitting of the light-hole and heavy-hole valence band at the H-point, $E_{LH-HH}$, are compared to the experimental values in Table \ref{tab:dftres}. Both, the gap and the valence band splitting are in very good agreement with the experiment, underestimating the experimental values slightly by $2\%$ and $1\%$, respectively. The complete band structure along high symmetry lines in the BZ is shown in Fig. \ref{bstr}.

From the wavefunctions, $\phi$, obtained from DFT, the transition dipole moments (TDMs) $\mathbf{d}^{nn^\prime}_{\mathbf{k}}$ between bands $n$ and $n^\prime$ at every k-point $\mathbf{k}$ are determined, 
\begin{equation}
\label{tdm}
 d^{nn^\prime}_{\mathbf{k}}  =  \frac{\hat{\mathbf{e}}}{\epsilon_{n\mathbf{k}} -\epsilon_{n'\mathbf{k}}} \cdot \left\langle \phi_{n'\mathbf{k}} \left|  \frac{\partial (\mathbf{ H} - \epsilon_{n\mathbf{k}} \mathbf{S})}{\partial \mathbf{k}} \right|  \phi_{n\mathbf{k}} \right\rangle \, .
\end{equation}
Here $\epsilon$ denotes the single particle energies and $\hat{\mathbf{e}}$ is the polarization direction. $\mathbf{ H}$ is the Hamilton operator for the cell periodic wavefunctions and $\mathbf{S}$ is the corresponding overloap operator\cite{Gajdos06}.
An overview of the TDMs projected onto the $z$-direction for optical fields polarized parallel to the $c$-axis ($E \parallel c$) and onto the  $x$-direction for $E \perp c$ is shown in Fig. \ref{dips}. Here, the modulus of the TDMs between the two lowest conduction bands and four highest valence bands are shown in a momentum vector plane spanned by the $\Gamma$-, A-, H-, L-, M-, and K-point of the BZ.

While the scale of the color map in Fig.~\ref{dips} is capped at 10, the maximum value of the dipoles between two valence bands or two conduction bands far exceeds that limit.
This can be explained from Eq. \ref{tdm}, since the bands of the same type are very close to each other up to the point of almost becoming degenerate, so that the factor $1/(\epsilon_{n\mathbf{k}} -\epsilon_{n'\mathbf{k}})$ becomes very large.
For the interband dipoles between a conduction and a valence band for $E\perp c$, the strongest dipole coupling is found around the direct bandgap at the H-point.
For $E\parallel c$ the interband dipoles involving the two highest valence bands are vanishingly small at the H-point, however,
they become stronger when moving away from the H-point along the H-L-line.
Only the interband dipoles of the two lower-lying valence bands, VB3 and VB4, are significant at the H-point. We will utilize this feature to simplify the optical response calculations for $E\parallel c$ by omitting the two upper valence bands, VB1 and VB2.

Generally, strong dipole coupling is found in parameter regions where the bands are close to each other. However, there are differences between the dipoles for $E\parallel c$ and $E\perp c$ although the band energies are the same for both polarization directions. E.g., the intraband dipoles are strong along the $\Gamma$-K-M line for $E\parallel c$, while there is no significant coupling  for $E\perp c$.
Conversely, the coupling along the H-K-line is much stronger for $E\perp c$ than for $E\parallel c$.

\section{Incoherent Resonant Nonlinearities}
\label{sec:abs}
In order to test the results of our DFT calculations we use the band structures, wavefunctions and TMDs to evaluate absorption spectra for Te and compare them to experimentally measured results.
The absorption is calculated for two polarization directions of the exciting light field, $E \parallel c$ and $E \perp c$.
In the BZ, these directions correspond to the H-K- and H-L-H-A-path, respectively.

Linear absorption spectra are computed by applying an arbitrarily small field $E(t)$ and calculating the material response $P(t)$ by solving the equations of motion for the microscopic polarizations, $p^{j i}_{\mathbf{k}}$, i.e. the SBE\cite{lindberg1988,girndt1997}:
\begin{align}
\label{sbe_eq}
 \frac{\mathrm{d}}{\mathrm{d}t} p^{j_1 i_1}_{\mathbf{k}} = & \frac{1}{i \hbar} ( \sum_{i_2, j_2}  \left[ \tilde{\epsilon}^{h}_{j_1 j_2,\mathbf{k}} \delta_{i_1 i_2} + \tilde{\epsilon}^{e}_{i_1 i_2, \mathbf{k}} \delta_{j_1 j_2} \right] p^{j_2 i_2}_{\mathbf{k}} \\
  \notag & \quad \quad \quad + \left[1 - f^{e}_{i_1, \mathbf{k}} - f^{h}_{j_1, \mathbf{k}} \right] \Omega^{i_1 j_1}_{\mathbf{k}} ) \\
  \notag &+ \left. \frac{\mathrm{d}}{\mathrm{d}t}  p^{j_1 i_1}_{\mathbf{k}} \right\vert_{\text{corr}}
\end{align}
with the renormalized electron and hole energies
\begin{align}
 \tilde{\epsilon}^{e}_{i_1 i_2, \mathbf{k}} &= \epsilon^{e}_{i_1,\mathbf{k}} \delta_{i_1 i_2} - \sum_{i_3, q} V^{i_1 i_3 i_2 i_3}_{\mathbf{k}-\mathbf{q}} f^{e}_{i_3, \mathbf{q}} \\
 \tilde{\epsilon}^{h}_{j_1 j_2, \mathbf{k}} &= \epsilon^{h}_{j_1,\mathbf{k}} \delta_{j_1 j_2} - \sum_{j_3, q} V^{j_2 j_3 j_1 j_3}_{\mathbf{k}-\mathbf{q}} f^{h}_{j_3, \mathbf{q}} \\
\end{align}
and the renormalized generalized Rabi frequency
\begin{align}
 \Omega^{i_1 j_1}_{\mathbf{k}} &= - d^{i_1 j_1}_\mathbf{k} E(t) - \sum_{i_2, j_2, q} V^{i_1 j_2 j_1 i_2}_{\mathbf{k}-\mathbf{q}} p^{j_2 i_2}_{\mathbf{q}} \, .
\end{align}
Here, $i_1, i_2, i_3$ are electron band indices and $j_1, j_2, j_3$ are hole band indices.
Like the dipole matrix elements, the Coulomb matrix elements $V$ are evaluated using the DFT wavefunctions.

For the linear absorption calculations, the material is assumed to be in the unexcited ground state and the field is too weak to create carriers such that the occupations for electrons/holes $f^{e/h}$ remain zero. For gain calculations, the carriers are assumed to be in thermal equilibrium and described by Fermi distributions within the respective bands. This fully microscopic approach has been shown to yield very good quantitative agreement with the experiment for a wide variety of materials spanning the mid-IR to visible wavelength ranges (see e.g. Ref.\onlinecite{nlcstr-web-page} for examples). 

The term $\left. \frac{\mathrm{d}}{\mathrm{d}t}  p^{j_1 i_1}_{\mathbf{k}} \right\vert_{\text{corr}}$ summarizes higher order correlations that include the electron-electron and electron-phonon scattering which lead to the dephasing of the polarization and the resulting homogeneous broadening of the spectra. We include the scatterings on a fully microscopic level by solving the corresponding quantum-Boltzman type scattering equations. Standard literature parameters are used for the phonon scattering as discussed in Ref.\onlinecite{girndt1997}. The explicit calculation of the dephasing processes not only eliminates adjustments requiring empirical parameters, but has also been shown essential to obtain correct lineshapes, amplitudes, spectral positions and density dependencies. 

From the Fourier transform of the macroscopic polarization $P(t) = \sum_{i,j,\mathbf{k}} p^{ji}_{\mathbf{k}} d^{ij*}_{\mathbf{k}}$, the absorption coefficient $\alpha$ is calculated according to
\begin{align}
 \alpha(\omega) = \frac{\omega}{\epsilon_0 n_r(\omega) c E(\omega)} \text{Im} \left[ P(\omega) \right] \, .
\end{align}

\begin{figure}[htbp]
    \includegraphics[width=0.4\textwidth]{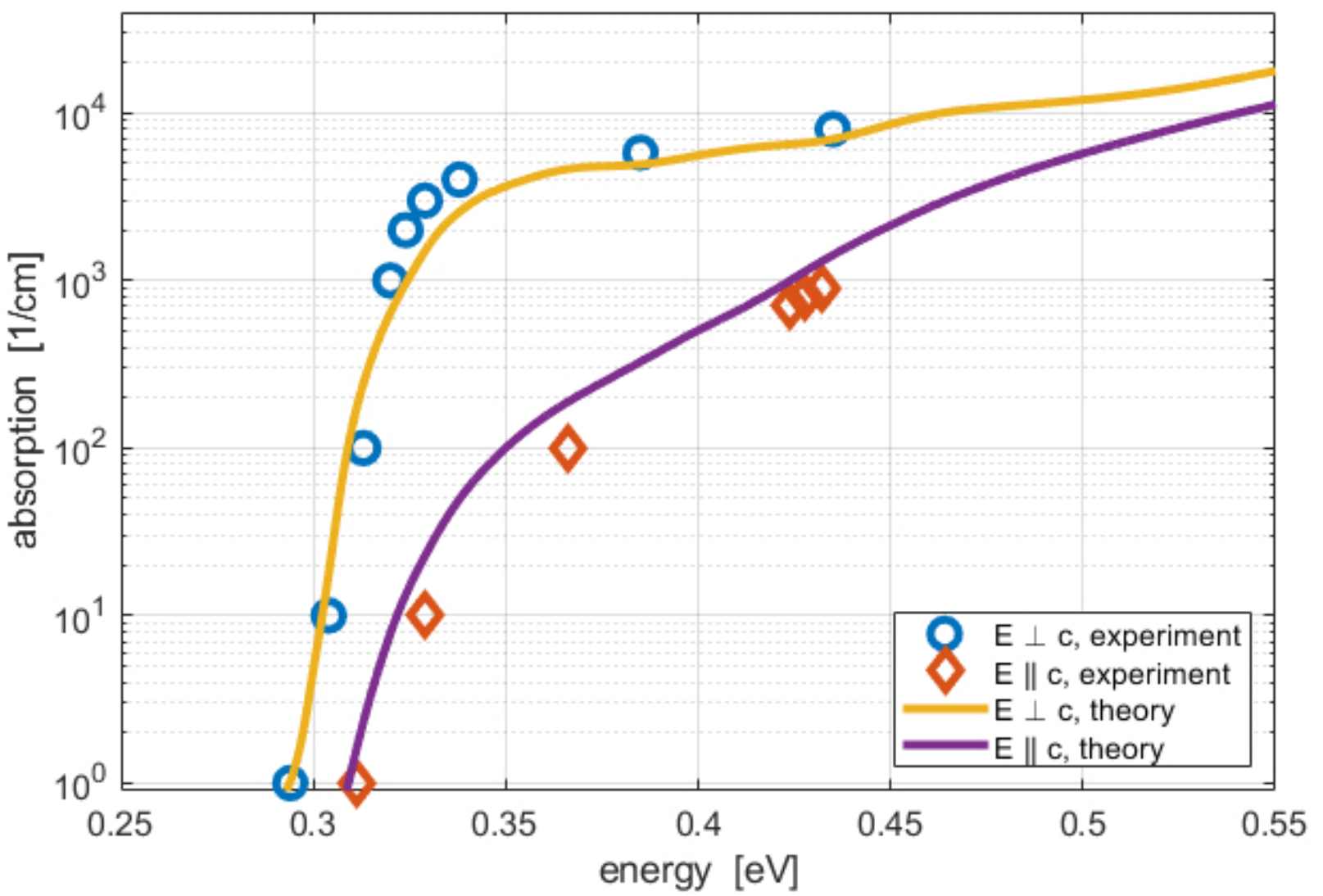}
    \caption{Room temperature material absorption of Te for light polarized $\parallel c$ (smaller) and $\perp c$ (larger). Solid lines: theoretical results based on DFT. Symbols: experimental data extracted from Ref.\onlinecite{Tutihasi1969}. The experimental data was shifted by $14\,\mathrm{meV}$ to lower energies.}
    \label{fig_abs}
\end{figure}
In Fig.~\ref{fig_abs}, we plot the resulting absorption spectra for the polarizations $E \perp c$ and $E \parallel c$. Especially near the bandgap, the absorption for $E \perp c$ is much larger than the one for $E \parallel c$ due to the weaker coupling between the topmost valence bands and the conduction bands near the bandgap for $E \parallel c$ (see Sec. \ref{dftres}). The blue and red dots in Fig.~\ref{fig_abs} show the results of measurements extracted from Ref.\onlinecite{Tutihasi1969}. As we can see, our computed results agree well with the experimentally measured spectra.  

As noted in Ref.\onlinecite{Tutihasi1969}, it is difficult to determine the reason for the strong polarization dependence of the absorption from measurement alone. While the authors of Ref.\onlinecite{Tutihasi1969} assumed that the absorption for $E\parallel c$ is suppressed due to an indirect gap, Refs.\onlinecite{Rigaux66, Grosse68} concluded that the real reasons are the selection rules leading to forbidden transitions for this configuration. This assumption is fully confirmed by our DFT calculations.

The authors of Ref.\onlinecite{Tutihasi1969} state the bandgap of their sample to be around $0.335-0.337\,\mathrm{eV}$ compared to our value of $0.323\,\mathrm{eV}$ and the value of about $0.33\,\mathrm{eV}$ from Refs.\onlinecite{Adenis1989, Anzin1977, Caldwell1959}. In Ref.\onlinecite{Choi19}, the authros report that it was possible to shift the bandedge photoluminescence of their sample by about $24\,\mathrm{eV}$ through annealing. This indicates that the bandgap of Te can vary due to  effects like sample quality or strain by amounts that can explain the difference found between our results and those in Ref.\onlinecite{Tutihasi1969}. In Fig.~\ref{fig_abs}, we account for the difference in bandgaps by shifting the experimental data extracted from Ref.\onlinecite{Tutihasi1969} by $14\,\mathrm{meV}$. 

\begin{figure}[htbp]
    \includegraphics[width=0.48\textwidth]{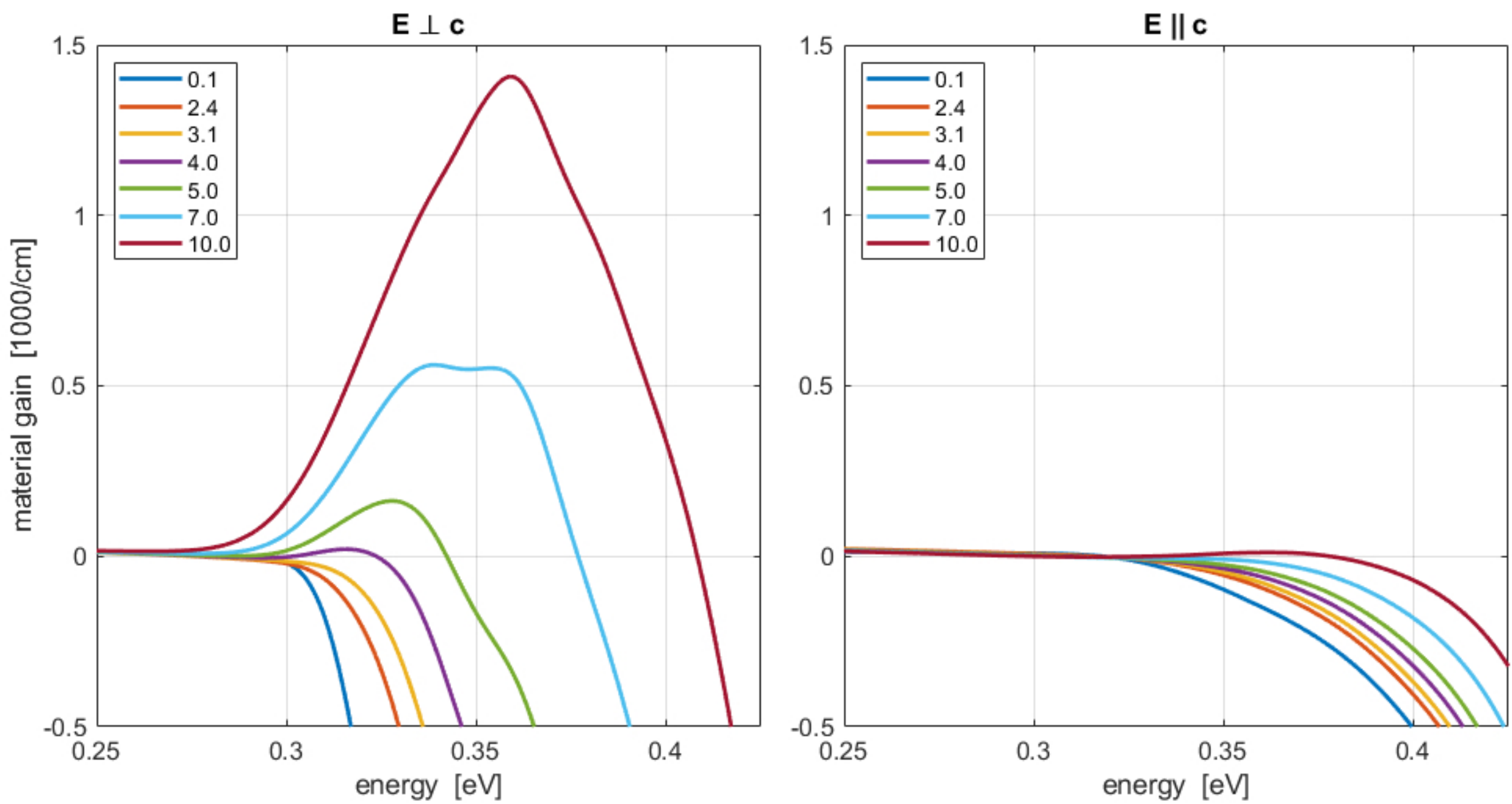}
    \caption{Room temperature material gain (negative absorption) spectra of Te for light polarized $\perp c$ (left) and $\parallel c$ (right) at various carrier densities. The carrier densities are given in the labels in units of $10^{19}/\mathrm{cm}^3$.}
    \label{fig_gain}
\end{figure}

Encouraged by the good agreement of the computed and measured linear absorption spectra, we use our microscopic approach to investigate the nonlinear optical properties of bulk Te. In a first step, we assume that the material has been excited to generate significant densities of incoherent electron and hole populations in the respective bands. As an example, we show in Fig.\ref{fig_gain} the calculated optical material gain ($-  \alpha(\omega)$) for $E \perp c$  and $E \parallel c$ and various carrier densities. We see that for $E\perp c$ gain begins to occur for carrier densities above $4\times 10^{19}/\mathrm{cm}^3$. For densities above about $7\times 10^{19}/\mathrm{cm}^3$ the peak gain shifts from the CB1-VB1 transition with a peak around $0.33-0.35\,\mathrm{eV}$ to the second conduction band transition, CB2-VB1, with a peak near $0.37\,\mathrm{eV}$.

As has been seen in the linear absorption spectra, the TDMs are much smaller for $E\parallel c$ than for $E\perp c$  in the spectral range where gain would occur. This leads to virtual no gain at all for this polarization direction at realistic carrier densities.

Assuming the same excitation conditions, the resulting photo luminescence (PL) is calculated by solving the SLE\cite{sle}, i.e., the microscopic equations of motion for the photon assisted polarizations. Structurally, the SLE have the same form as the SBE, Eq.(\ref{sbe_eq}), but include higher excitonic correlations as an additional source term. As for the SBE, we include in our SLE evaluations the electron-electron and electron-phonon scattering on a fully microscopic level.

\begin{figure}[htbp]
    \includegraphics[width=0.48\textwidth]{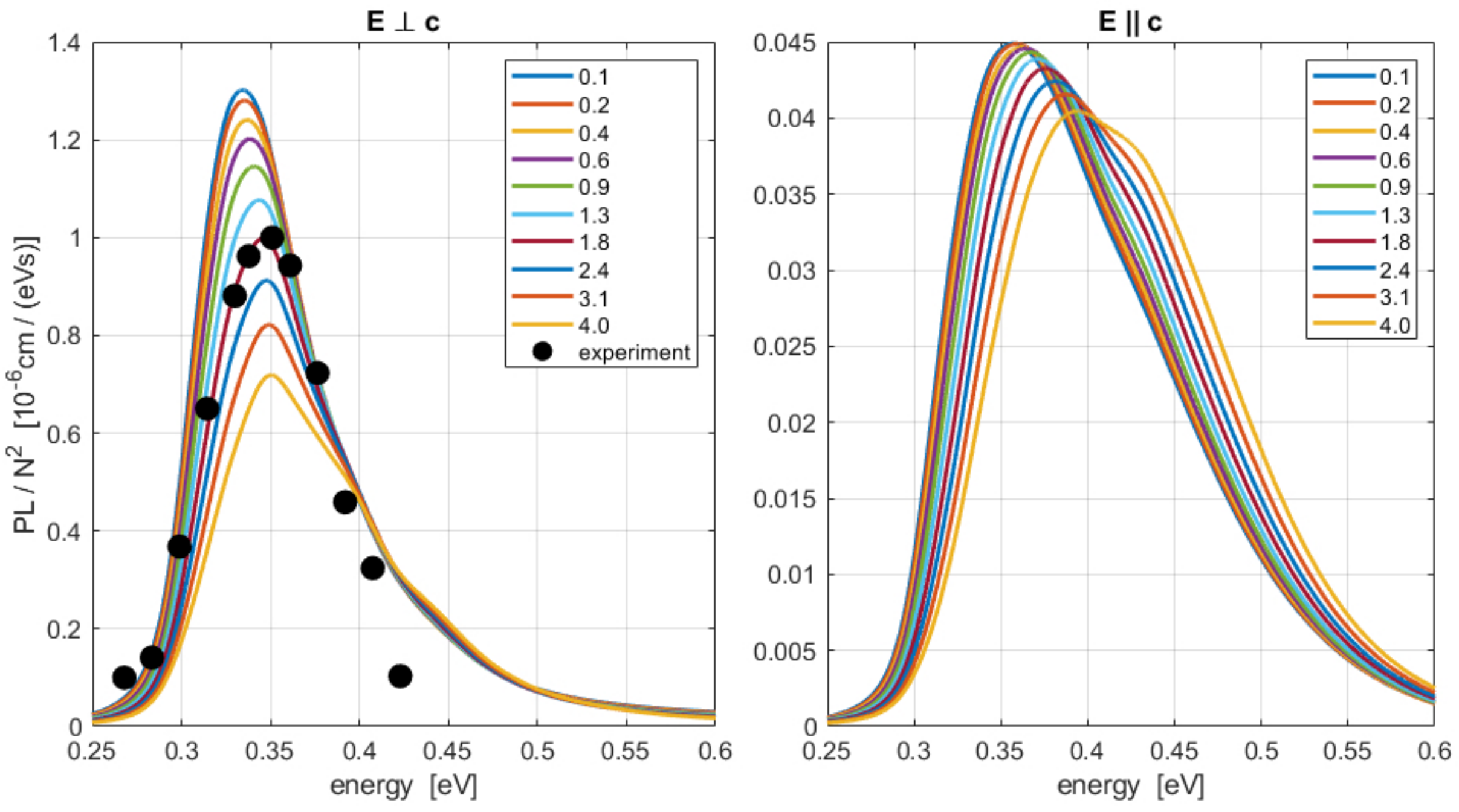}
    \caption{Theoretrical (lines) and experimental (symbols) room temperature photo luminescence spectra of Te for light polarized $\perp c$ (left) and $\parallel c$ (right) at various carrier densities. The theoretical spectra have been divided by the respective density squared. The carrier densities are given in the labels in units of $10^{19}/\mathrm{cm}^3$. The experimental data extracted from Ref.\onlinecite{Choi19} are given in arbitrary units.}
    \label{fig_pl}
\end{figure}

Fig.\ref{fig_pl} shows PL spectra for $E\perp c$ and $E\parallel c$ at various levels of electron-hole-pair populations. In the low density regime, the PL scales quadratically with the carrier density. Plotting the PL divided by the square of the density as in Fig.\ref{fig_pl} reveals deviations for higher excitation levels from this quadratic variation that are due to phase space filling\cite{apl-phase-space05}. In this regime, the density dependence becomes less than quadratic and the PL peak shifts to higher transition energies. For the case of $E\parallel c$ the peak shift is stronger and the amplitude reduction is slower. These features can be attributed to the fact that the TDMs for $E\parallel c$ increase significantly with increasing energy above the gap which enhances energetically higher PL contributions. Like the gain and absorption, the PL is much weaker for $E\parallel c$ than for $E\perp c$ due to the much smaller TDMs in the energy region of interest. This agrees with the experimentally observed dominant polarization $E\perp c$ of PL in Ref.\onlinecite{Benoit65}. The spectral position as well as the lineshape of our calculated PL agrees very well with experimentally measured data from Ref.\onlinecite{Choi19} that we include in Fig.\ref{fig_pl} for comparison. This demonstrates the  high accuracy of the fully microscopic modelling approach including the explicit treatment of scattering processes that lead to an almost perfect agreement with the experimentally observed linewidth of about $80\,\mathrm{meV}$.

\section{Coherent Off-Resonant Nonlinear Response}

\subsection{Microscopic Approach}

In order to model the nonlinear optical response of a crystal to a strong exciting THz field, the coupled dynamics of interband polarizations and intraband currents have to be investigated. For this purpose, we again use the SBE. However, in contrast to the quasi stationary nonlinear response investigated so far, we now have to explicitly include the nonequilibrium carrier dynamics. In particular, the strong long-wavelength excitation field leads to an acceleration of carriers along the bands throughout the entire BZ. Thus, the results depend critically on the dispersion relation across the whole BZ. Furthermore, pulse propagation effects have to be included in order to study the dependence of HHG on sample length.

In earlier studies, we have shown that for the strongly off-resonant excitation assumed here, the Coulomb renormalizations have a negligible inlfuence\cite{huttner-even-hhg17} such that the equations of motion can be simplified to 
\begin{align} \label{eq:sbep}
 i \hbar \frac{\mathrm{d}}{\mathrm{d} t} p^{\mathrm{h}_i \mathrm{e}_j}_{\mathbf{k}} &= \left( \epsilon^{\mathrm{e}_j}_{\mathbf{k}} + \epsilon^{\mathrm{h}_i}_{\mathbf{k}} + i \lvert e \rvert E_\text{THz} (t) \nabla_{\mathbf{k}} \right) p^{\mathrm{h}_i \mathrm{e}_j}_{\mathbf{k}} \\
 \notag &- \hbar \Omega^{\mathrm{h}_i \mathrm{e}_j}_{\mathbf{k}} (t) \left( 1 - f^{\mathrm{e}_j}_{\mathbf{k}} - f^{\mathrm{h}_i}_{\mathbf{k}} \right) + \Gamma^{\mathrm{h}_i \mathrm{e}_j}_{\mathbf{k}} \\
 \notag &+ \sum_{\mathrm{e}_\lambda \neq \mathrm{e}_j} \left[ \hbar \Omega^{\mathrm{h}_i \mathrm{e}_\lambda}_{\mathbf{k}} (t) p^{\mathrm{e}_\lambda \mathrm{e}_j}_{\mathbf{k}} - \hbar \Omega^{\mathrm{e}_\lambda \mathrm{e}_j}_{\mathbf{k}} (t) p^{\mathrm{h}_i \mathrm{e}_\lambda}_{\mathbf{k}} \right] \\
 \notag &+ \sum_{\mathrm{h}_\lambda \neq \mathrm{h}_i} \left[ \hbar \Omega^{\mathrm{h}_i \mathrm{h}_\lambda}_{\mathbf{k}} (t) p^{\mathrm{h}_\lambda \mathrm{e}_j}_{\mathbf{k}} - \hbar \Omega^{\mathrm{h}_\lambda \mathrm{e}_j}_{\mathbf{k}} (t) p^{\mathrm{h}_i \mathrm{h}_\lambda}_{\mathbf{k}} \right]\\
\notag &+ \left. \frac{\mathrm{d}}{\mathrm{d}t}  p^{\mathrm{h}_i \mathrm{e}_j}_{\mathbf{k}} \right\vert_{\text{corr}}
\end{align}
\begin{align}
 \hbar \frac{\mathrm{d}}{\mathrm{d} t} f^{\mathrm{e}_i}_{\mathbf{k}} &= - 2 \hbar \; \times \\
 \notag  \times \;  &\text{Im} \left[ \sum_{\mathrm{e}_\lambda \neq \mathrm{e}_i} \Omega^{\mathrm{e}_\lambda \mathrm{e}_i}_{\mathbf{k}} (t) \left( p^{\mathrm{e}_\lambda \mathrm{e}_i}_{\mathbf{k}} \right)^* + \sum_{\mathrm{h}_\lambda} \Omega^{\mathrm{h}_\lambda \mathrm{e}_i}_{\mathbf{k}} (t) \left( p^{\mathrm{h}_\lambda \mathrm{e}_i}_{k} \right)^* \right] \\
 \notag &+ \lvert e \rvert E_\text{THz} (t) \nabla_{\mathbf{k}} f^{\mathrm{e}_i}_{\mathbf{k}} + \Gamma^{\mathrm{e}_i}_{\mathbf{k}} .
\end{align}
We have similar expressions for the intraband polarizations between conduction bands $p^{\mathrm{e}_i \mathrm{e}_j}_{\mathbf{k}}$ and between valence bands $p^{\mathrm{h}_i \mathrm{h}_j}_{\mathbf{k}}$ and the carrier occupations in the valence band $f^{\mathrm{h}_i}_{\mathbf{k}}$, respectively. For HHG, we model the dephasing of the polarization as represented by the last term in Eq.(\ref{eq:sbep}) using a dephasing time $T_2=40\,$fs.

The macroscopic polarization $P(t) = \sum_{\lambda, \lambda^\prime, \mathbf{k}} d^{\lambda \lambda^\prime}_{\mathbf{k}} p^{\lambda \lambda^\prime}_{\mathbf{k}}$ and the macroscopic current $J(t) = \sum_{\lambda, \mathbf{k}} j_\lambda (\mathbf{k}) f^{\lambda}_{\mathbf{k}}$ due to the acceleration of carriers along the bands contribute to the emitted electric field $E_{\text{out}} (t) \propto \frac{\partial}{\partial t} P(t) + J(t)$ and create the characteristic local high harmonic emission spectrum which is given by the emission intensity $I_\text{out} (\omega) \propto \lvert \omega P (\omega) + i J (\omega) \rvert^2$.

In order to gain some insights before doing the full propagation calculations, we performed local evaluations which need significantly less numerical effort. Here, we use a one-dimensional $\mathbf{k}$-space model which assumes that carriers are predominantly excited near the fundamental gap, i.e. near the $H$-point with negligible momentum perpendicular to the field. For linearly polarized light, the carriers are then driven along a one dimensional path through the BZ. For $E\parallel c$ the path is from $K$ to $H$ and back to $K$. For $E\perp c$ the path goes from $A$ to $H$ to $L$ and back. For all HHG simulations we assume excitation with a Gaussian pulse, $E(t)=E_0\exp{-(t/\sigma)^2}cos(\omega_0t)$, with a width $\sigma=100\,$fs  and a central frequency $\omega_0$ corresponding to a wavelength of $10.6\,\mu$m.

In a first step, we use this local model to identify those bands that are relevant for HHG generation under typical off-resonant excitation conditions. Clearly, the HHG signal is dominated by transitions between those bands which are energetically closest to the bandgap unless these transitions are suppressed due to symmetry selection rules leading to small TDMs.
As can be seen in Fig.\ref{bstr}, only four valence and two electron bands are in the energetically relevant region. Since the TDMs presented in Fig.~\ref{dips} show that the coupling of the top two valence bands to the lowest two conduction bands vanishes at the H-point for $E \parallel c$, we studied whether these bands are significant for the resulting HHG spectrum. 

A comparison of the computed spectra including different valence bands is shown in Fig.~\ref{fig:model} a). We note that by considering only the bottom two valence bands, we obtain a spectrum that agrees rather well with the full six-band calculation, allowing us to reduce the complexity of our propagation studies for $E \parallel c$ by including only this subset of bands. In contrast, for the $E \perp c$ configuration, the top two valence bands dominate the response and are thus included in the HHG simulations.

\subsection{Phase of Transition Dipole Matrix Elements}

In general, the TDMs presented in Sec.~\ref{dftres} are complex valued.
To illustrate the influence of the phases on the HHG emission, we consider a perturbative power series of the polarization response to an electric field for a situation with two valence bands $h1, h2$ and one conduction band $e$.  In first order of the field, all polarizations and occupations are $0$, so that we obtain from Eq.~\ref{eq:sbep} 
\begin{align}
 \left( p^{\mathrm{h}_1 \mathrm{e}}_{\mathbf{k}} (t) \right)^{(1)} \propto \frac{1}{\hbar \omega} d^{\mathrm{e} \mathrm{h}_1}_{\mathbf{k}} E(t) \, .
\end{align}
The resulting macroscopic polarization then yields
\begin{align}
 \left( P_{\mathrm{h}_1 \mathrm{e}} (t) \right)^{(1)} = \sum_\mathbf{k} d^{\mathrm{h}_1 \mathrm{e}}_{\mathbf{k}} \left( p^{\mathrm{h}_1 \mathrm{e}}_{\mathbf{k}} (t) \right)^{(1)} \propto \frac{\lvert d^{\mathrm{h}_1 \mathrm{e}}_{\mathbf{k}} \rvert^2}{\hbar \omega} E(t) \, .
\end{align}
Hence, the phase of the TDMs in this first-order response is irrelevant.
However, since the polarizations are non-zero in second order, the creation of a polarization between the valence bands allows for an indirect excitation into the conduction band, 
\begin{align}
 \left( p^{\mathrm{h}_1 \mathrm{e}}_{\mathbf{k}} (t) \right)^{(2)} \propto \frac{d^{\mathrm{e} \mathrm{h}_1}_{\mathbf{k}}}{\hbar \omega} E(t) + \frac{d^{\mathrm{h}_2 \mathrm{h}_1}_{\mathbf{k}} d^{\mathrm{e} \mathrm{h}_2}_{\mathbf{k}}}{2 \hbar^2 \omega^2} E^2 (t) + ...
\end{align}
This leads to a term in the macroscopic polarization
\begin{align}
\label{eq_ppp}
 \left( P_{\mathrm{h}_1 \mathrm{e}} (t) \right)^{(2)} \propto \sum_\mathbf{k} \frac{d^{\mathrm{h}_2 \mathrm{h}_1}_{\mathbf{k}} d^{\mathrm{h}_1 \mathrm{e}}_{\mathbf{k}} d^{\mathrm{e} \mathrm{h}_2}_{\mathbf{k}}}{2 \hbar^2 \omega^2} E^2 (t) + ...
\end{align}
where the phases of the TDMs do not vanish. If, e.g., one of the TDMs in Eq.(\ref{eq_ppp}) is antisymmetric in $\mathbf{k}$ and the other two are symmetric, the integration over $\mathbf{k}$ will lead to a zero contribution to the macroscopic polarization and resulting HHG signal while a strong non-zero contribution would be obtained if the phases are neglected. Thus, the phases need to be considered correctly in order to obtain the correct symmetry-related selection rules and amplitudes in the HHG calculations. 

It was shown in Ref. \onlinecite{huttner-even-hhg17} that quantum interference between intraband and interband polarizations can lead to the appearance of even harmonics. Moreover, if one neglects the phases of the TDMs, even harmonics would be allowed for all systems with three or more bands - which is known not to be the case. Thus, the correct inclusion of the phases is essential to obtain the correct selection rules for HHG.

While the TDMs are complex valued, the plot in Fig.~\ref{dips} only shows the absolute value.
In DFT, the Schr\"odinger equation of every $\mathbf{k}$-point is solved individually, so that there is no phase relation between different $\mathbf{k}$-points.
Therefore, the computed TDMs contain a random phase which is not smooth across the BZ.
As it turns out, this random phase can be eliminated for all $\mathbf{k}$-points by evaluating the product of the three complex TDMs connecting the bands $n$, $n^\prime$ and $n^{\prime \prime}$ in a circular way, e.g. $T^{n n^{\prime} n^{\prime \prime}}_{\mathbf{k}} = d^{n n^{\prime}}_{\mathbf{k}} d^{n^{\prime} n^{\prime \prime}}_{\mathbf{k}}  d^{n^{\prime \prime} n}_{\mathbf{k}}$.
The random phase of each band vanishes in the product, so that the phase of $T^{n n^{\prime} n^{\prime \prime}}_\mathbf{k}$ along any direction in the BZ is smooth.
Since this only gives us the phase information about the product of three TDMs, this phase is applied to one of the constituent TDMs while taking the other ones as purely real.
In that way, the triple product will have the correct phase.

\begin{figure}[htbp]
    \includegraphics[width=0.48\textwidth]{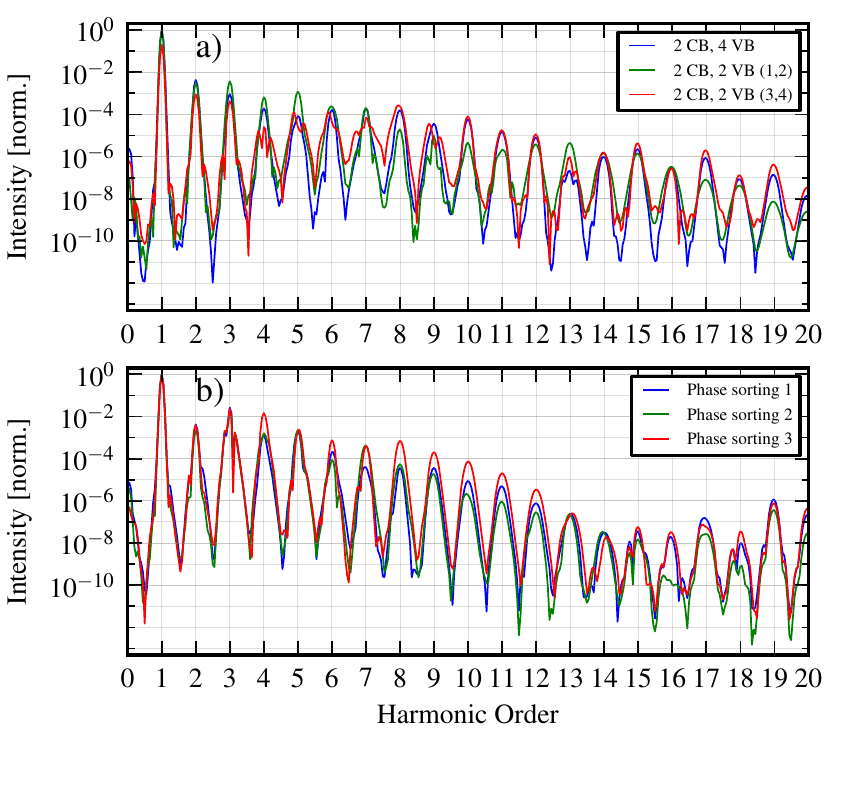}
    \caption{Polarization part of $E \parallel c$ HHG emission in Te. (a) Influence of different choices of bands on HHG emission. (b) Influence of choice of TDM phases on HHG emission.}
    \label{fig:model}
\end{figure}

\begin{figure}[htbp]
    \includegraphics[width=0.48\textwidth]{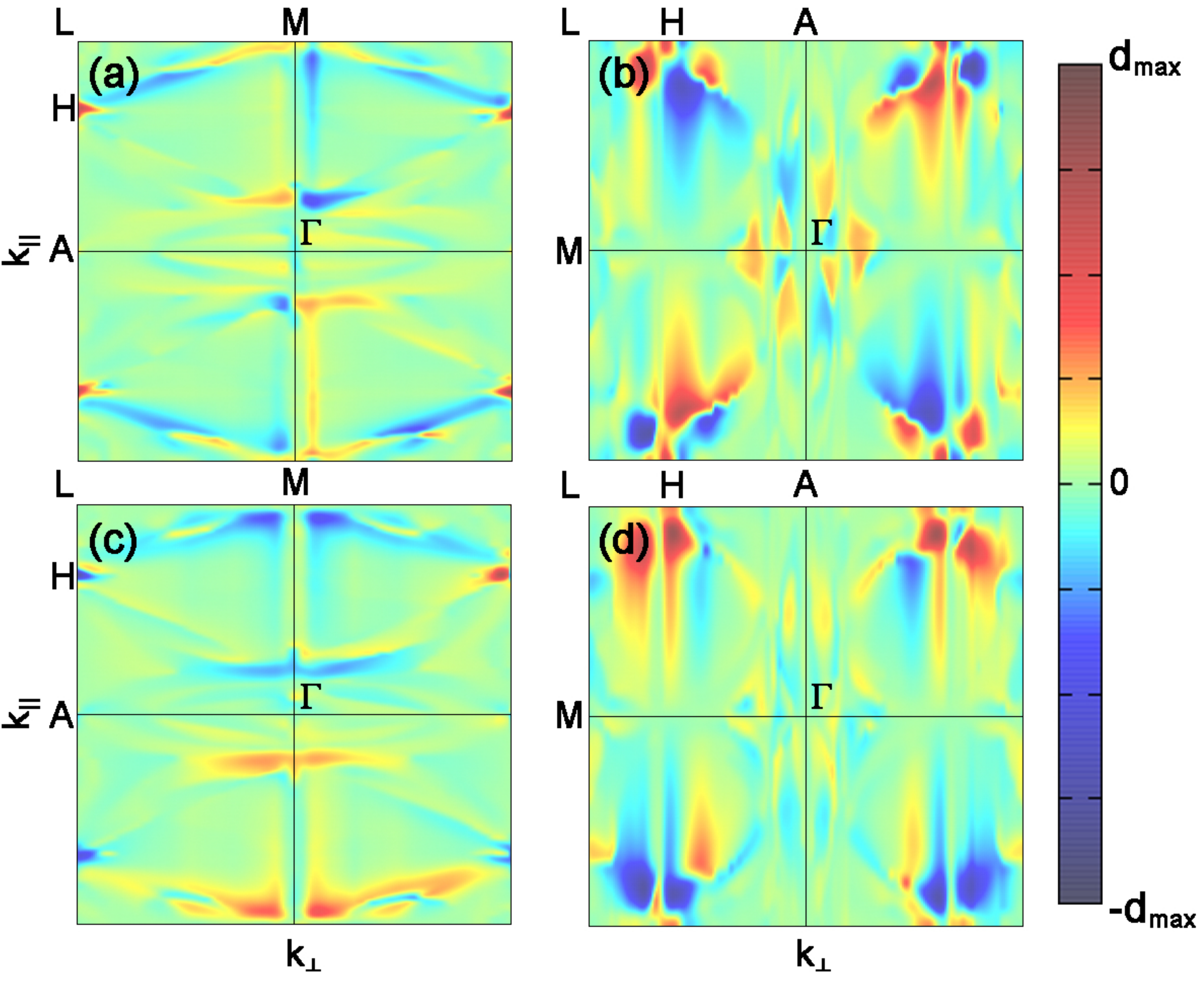}
    \caption{Complex dipole matrix elements between the lowest conduction and highest valence band. (a) and (c) are the real and  imaginary parts for $E\perp c$. (b) and (d) are the real and imaginary parts for $E\parallel c$. $k_\perp$ ($k_\parallel$) is the momentum perpendicular (parallel) to the field polarization. $d_{max}= 6,\,8,\,4,\,$ and $5$ for (a), (b), (c), and (d), respectively.}
   \label{dip_e1h1}
\end{figure}

As an example, we show in Fig.\ref{dip_e1h1}, the complex TDMs for the respective transitions between the lowest electron and highest hole bands taken into account for $E \parallel c$ and $E \perp c$. In all four plots, the momentum parallel to the field polarization is vertically aligned. Once the phases of the dipoles are taken into account it becomes obvious that the Te system does not have pure radial or inversion symmetry. Thus, the $\mathbf{k}$-domain has to be expanded from the positive sector $\Gamma$-$A$-$L$-$M$ to four times the size to include also negative $k_x$ and $k_z$. For the $e1-h1$ transitions presented in Fig.\ref{dip_e1h1}, the real and imaginary parts of the dipoles for $E\parallel c$, shown in the two right-hand plots, appear nearly antisymmetric along the polarization direction. In contrast, the symmetry properties of the real and imaginary parts of the dipoles for $E\perp c$ are a little more ambiguous. As for the Te crystal itself, the TDMs do not have perfect (anti-) symmetry. This can be seen e.g. in the real parts of the TDMs for $E \perp c$ in Fig.\ref{dip_e1h1}. These are nearly symmetric near $H$ while they appear mostly antisymmetric in most regions of small $k_\perp$. The imaginary parts for $E\perp c$ in Fig.\ref{dip_e1h1}(c) appear mostly antisymmetric, with slight deviations around the H-point.

In our procedure to assign the TDM phase, we arbitrarily choose the dipoles onto which we impose the smoothed phase of the triple dipole products. In order to check how this choice influences the HHG spectrum, we calculated the polarization part of the spectra for different phase projections. As can be seen in Fig.~\ref{fig:model} b), our phase assignment does not influence the overall structure of the spectra, leading only to insignificant amplitude changes, so that the comparisons between HHG calculations for different intensities, propagation lengths etc. is robust against this choice for the dipole phases.

\subsection{High Harmonics in Te}

\begin{figure}[htbp]
    \includegraphics[width=0.48\textwidth]{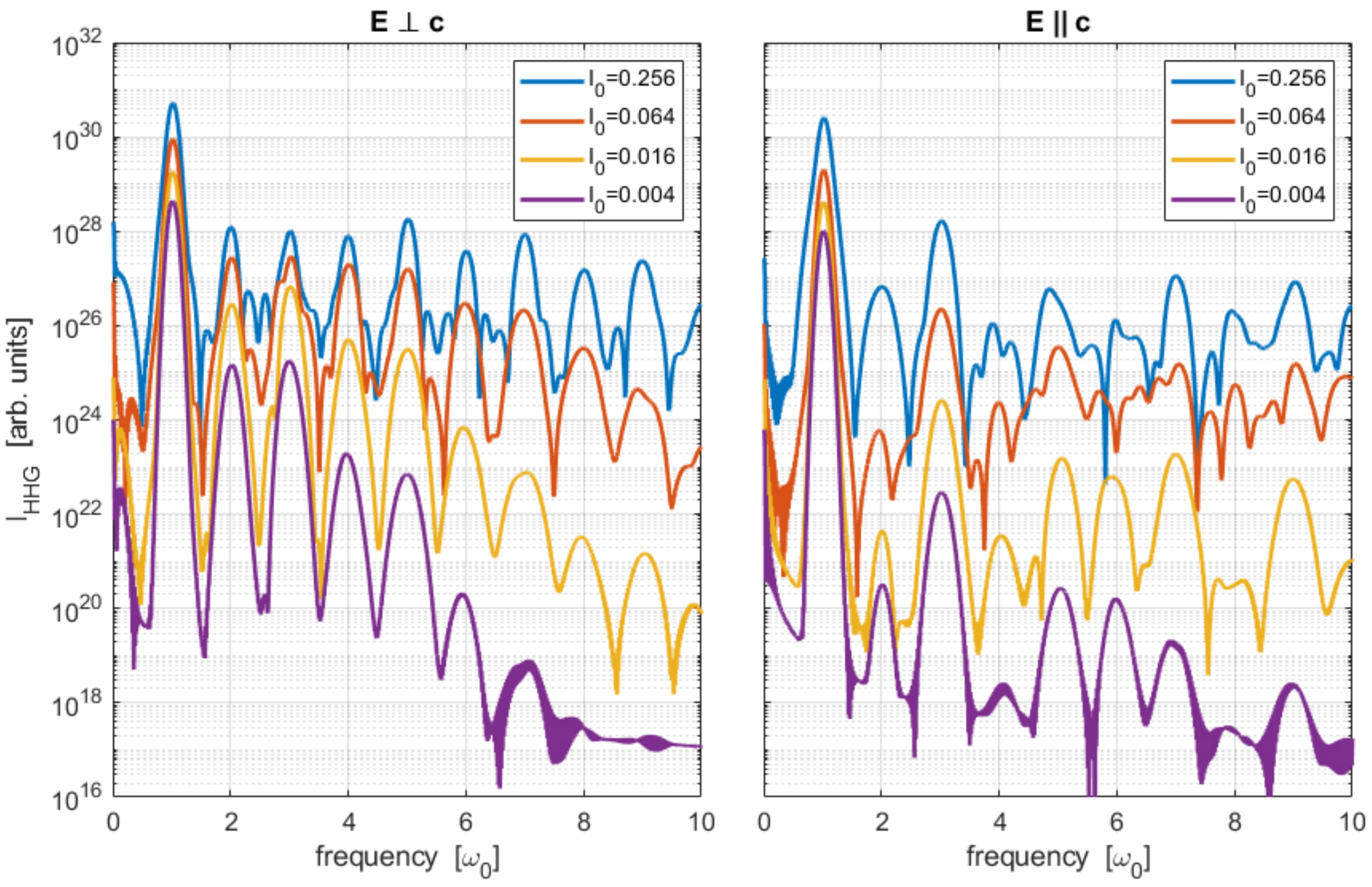}
    \caption{HHG spectra in Te for $E\perp c$ (left) and $E\parallel c$ (right) and various peak intensities $I_0$. Intensities given in the labels are in units of $10^{14}W/m^2$.}
    \label{hhg_ahl_hk_local}
\end{figure}

In order to determine the dependence of HHG production in Te on the field strength, we performed calculations for the material response only, without pulse propagation. Figure \ref{hhg_ahl_hk_local} shows the resulting emission for various intensities of the exciting pulse. 

For both polarization configurations, a significant signal above the bandgap (frequencies above the third harmonic) develops for peak intensities above about $10^{11}W/m^2$. A plateau starts to form for about 100 times higher intensities. Harmonics below the bandgap emerge rather quickly for $E\perp c$ and start to saturate already at amplitudes about three orders below that of the fundamental. For $E\parallel c$, the signal below the bandgap develops much slower with field intensity. In particular, the third harmonic shows less saturation for the intensities investigated here. The differences at and below the bandgap are due to the fact that the interband coupling is much weaker at and near the gap as can be seen from the absorption spectra.

Even harmonics are strongly suppressed for $E\parallel c$ while for $E\perp c$ no obvious discrimination occurs between even and odd harmonics. This behavior is a consequence of the symmetry of the dipole matrix elements. As in the case for the lowest electron-hole transition shown in Fig.\ref{dip_e1h1}, all dipoles that are relevant for even harmonics are nearly inversion symmetric for $E\parallel c$. This leads to a destructive quantum interference that suppresses the even harmonics. In contrast, for $E\perp c$ the relevant dipoles are dominantly symmetric which effectively enables quantum interference and allows for the even harmonics to reach similar levels as the odd harmonics.

\begin{figure}[htbp]
    \includegraphics[width=0.48\textwidth]{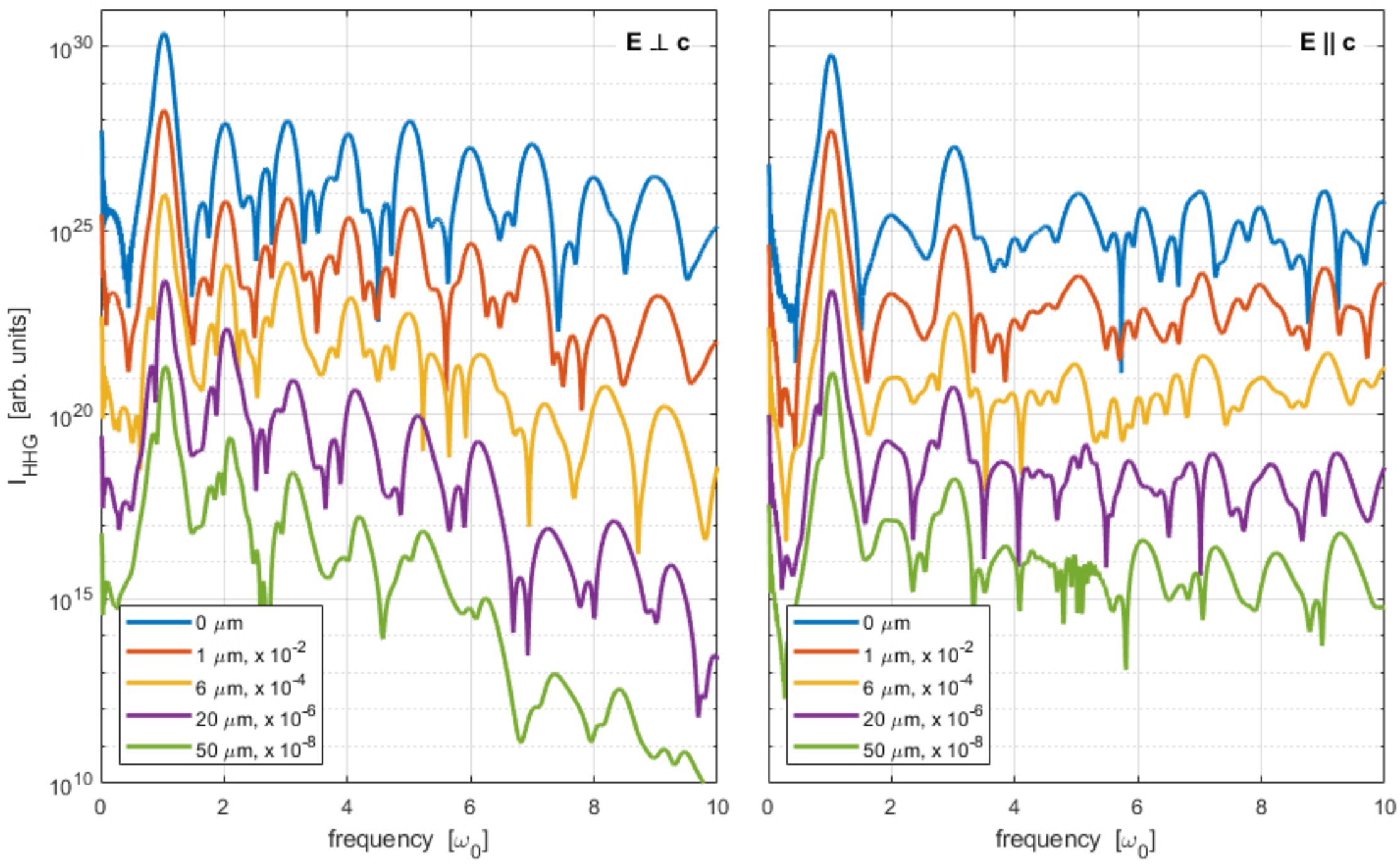}
    \caption{HHG spectra in Te for $E\perp c$ (left) and $E\parallel c$ (right), a peak pulse intensity of $0.128 \times 10^{14}W/m^2$ and for various propagation distances. Spectra for different propagation distances have been scaled by factors of 100 for better visibility. }
    \label{hhg_ahl_hk_prop}
\end{figure}

To evaluated HHG for samples of different thicknesses, we include pulse propagation effects by coupling the SBE (Eq.\ref{eq:sbep}) to a uni-directional pulse propagation solver as described in Ref. \cite{prl-pd} and references therein. As an example of the results, Fig.\ref{hhg_ahl_hk_prop} shows HHG spectra after propagation through Te for various distances. The initial pulse has a peak intensity of $0.128 \times 10^{14} W/m^2$.  

For $E\perp c$ the higher harmonics quickly weaken with propagation distance. In part this is a consequence of the gradual decreasing excitation pulse due to HHG and absorption of spectral components above the bandgap. In part this is also due to propagation induced dephasing \cite{prl-pd}. This weakening is less pronounced for $E\perp c$ since the absorption is weaker and less HHG signal is produced. Over the maximum propagation distance investigated here ($50\,\mu\mathrm{m}$) the amplitude of the fundamental drops by about a factor of ten for $E\perp c$ and only a factor of four for $E\parallel c$. The reduced amount of even harmonics for $E\parallel c$ likely also leads to a reduced amount of quantum interference and resulting  propagation induced dephasing within the remaining signal.

\section{Summary and Outlook}
In summary, we present a comprehensive microscopic analysis of optical nonlinearities in bulk Te. We determine the bandstructure, the optical dipoles, and the Coulomb interaction matrix elements using an DFT based approach. Investigating the near bandgap optical response for different levels of electron-hole-pair excitations, we numerically solve the stationary SBE and SLE to computed the strongly orientation dependent absorption and PL modifications. Comparing the linear absorption and PL spectra with experimental findings, we obtain excellent agreement. For elevated excitation levels, we obtain a transition from absorption to optical gain for $E\perp c$ gain with a peak in the technologically interesting mid-IR region. Since the TDMs are much smaller for $E\parallel c$ than for $E\perp c$  virtually no gain occurs for this polarization direction at realistic carrier densities. 

The generation of high-harmonic emission in Te is analyzed using the fully dynamic SBE systematically treating the nonequilibrium dynamics of the optically induced polarizations and currents. Pulse propagation effects are modeled by coupling the SBE to a unidirectional propagation solver that allows us to study the sample length and field orientation dependence of the even- and odd-order HHG for the different field polarization configurations. The importance of a correct treatment of the complex phases of dipole matrix elements for the correct description of optical selection rules is demonstrated.

As a next step, we plan to evaluate the intrinsic losses in bulk Te, in particular the Auger losses that typically hamper the laser application potential of mid-IR emitting structures. Furthermore, we will extend our comprehensive microscopic approach to low dimensional Te
\cite{Te2d} to investigate its nonlinear opto-electronic properties and device application potential. 

\begin{acknowledgments}
The authors thank D. Matteo, S. Tochitsky, UCLA, for stimulating discussions during the early part of these investigations and I. Kilen, M. Kolesic, University of Arizona, for development of the numerical HHG propagation code. The Marburg work was supported by the Deutsche Forschungsgemeinschaft (DFG) in the framework of the Research Training Group ``Functionalization of Semiconductors'' (GRK~1782) and the Collaborative Research Center SFB 1083. The authors thank the HRZ Marburg and CSC-Goethe-HLR Frankfurt for computational resources. The Tucson work was supported by the Air Force Office of Scientific Research under award number FA9550-17-1-0246.
\end{acknowledgments} 

\section*{Data Availability}
The data that support the findings of this study are available from the corresponding author upon reasonable request.

\end{document}